\documentclass{article}    

\usepackage{graphicx}

\title{\textbf{Quantum Brain oRules}}  
\author{Richard Mould\footnote{Department of Physics and Astronomy, State University of New York, Stony Brook,
\mbox{New York} 11794-3800; http://ms.cc.sunysb.edu/\~{}rmould}}  
\date{}    
   
\begin{document}             

\maketitle              

\begin{abstract}

Quantum mechanics traditionally places the observer `outside' of the system being studied and employs the Born interpretation. 
In this and related papers the observer is placed `inside' the system.  To accomplish this, special rules are required to engage
and interpret the Schr\"{o}dinger solutions in individual measurements.  The rules in this paper (called the oRules) do not
include the Born rule that connects probability with square modulus. 

It is required that the rules allow conscious observers to exist inside the system without empirical ambiguity -- reflecting our
own unambiguous experience in the universe.  This requirement is satisfied by the oRules. These rules are restricted to observer
measurements, so state reduction can only occur when an observer is present.

\end{abstract}

\section*{Introduction}
The method of this paper differs from that of traditional quantum mechanics in that it sees the observer in an ontological rather
than an epistemological context.  Traditional or standard quantum theory (i.e., Copenhagen) places the observer outside of the
system where operators and/or operations are used to obtain information about the system.    This is the epistemological model
shown in fig.\ 1.  The observer cannot here make continuous contact with the system \mbox{-- only} instantaneous contact.
\begin{figure}[t]
\centering
\includegraphics[scale=0.8]{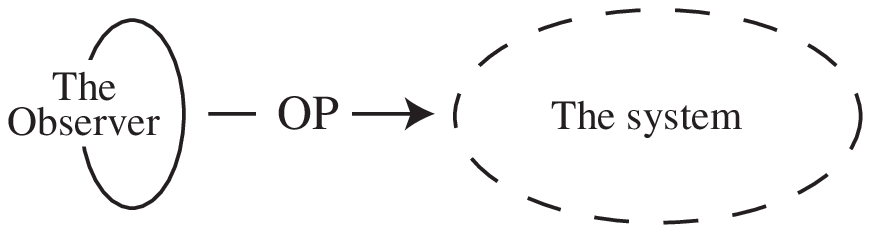}
\center{Figure 1: Epistemological Model (Copenhagen)}
\end{figure}

The large OP in fig.\ 1 might be a mathematical `operator' or a corresponding physical `operation'.  The observer makes a
measurement by choosing a formal operator that is associated with a chosen laboratory operation.  As a result, the observer is
forever outside of the observed system -- making operational choices. The observer is forced to act apart from the system as one
who poses theoretical and experimental questions to the system.  This model is both useful and epistemologically sound.  

	However, the special rules developed in this paper apply to the system by itself, independent of the possibility that an observer
may be inside, and disregarding everything on the outside.  This is the ontological model shown in \mbox{fig. 2}.
\begin{figure}[h]
\centering
\includegraphics[scale=0.8]{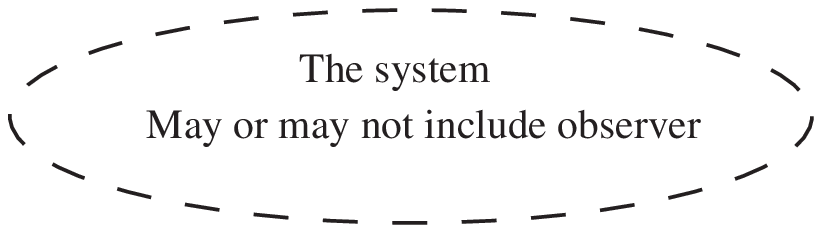}
\center{Figure 2: Ontological Model (requires special rules)}
\end{figure}

A measurement occurring inside this system is not represented by a formal operator.  Rather, it is represented by a physical
device that is itself part of the system.  If the sub-system being measured is $S$ and a detector is $D$, then a measurement
interaction is given by the entanglement $\Phi = SD$.  If an observer joins the system in order to look at the detector, then the
system state becomes $\Phi = SDB$, where $B$ is the brain state of the observer.  Contact between the observer and the observed is
continuous in this case.

The ontological model is able to place the observer inside the universe of things and give a full account of his conscious
experience there.  It is a more realistic view of the relationship between the observer and the rest of the universe, inasmuch as
a conscious observer is always `in principle' includable in a wider system. The ontological model is a departure from the
traditional theory and has three defining characteristics:     (1) It includes observations given by $\Phi = SDB$, (2) it allows
conscious observations to be continuous, and (3) it rejects the Born rule.  In place of the Born rule, special rules like the
oRules of this paper allow physics to unambiguously predict the continuous experience of an observer in the system.

\vspace{.4cm}

Quantum mechanical measurement is sometimes said to refer to ensembles of observations but not to individual observations.  In
this paper we propose a set of four special rules that apply to individual measurements in the ontological model.  They are called
\emph{oRules (1-4)}, and do not include the long-standing Born interpretation of quantum mechanics.  Instead, probability is
introduced (only) through the notion of \emph{probability current}.  Furthermore, these rules describe state reductions (i.e.,
stochastic reductions or collapses) that are associated with an `observer' type measurement -- that is, they occur only in the
presence of an observer.  To this extent, they reflect the  views of Wigner \cite{EW} and von Neumann \cite{JvN}.  The
oRules are demonstrated below in several different physical situations.  I claim that they are a consistent and complete set of
rules that are capable of giving an ontological description of any individual measurement or interaction in quantum mechanics.

The oRules (1-4) are also to be found under the name ``rules (1-4)" in an earlier paper \cite{RM1}, where they are developed
somewhat differently. 

These rules are not themselves a formal theory of measurement.  I make no attempt to understand \emph{why} they work, but strive
only to insure that they do work. Presumably, a formal theory can one day be found to explain these rules in the same way that
atomic theory explains the empirically discovered rules of atomic spectra, or in the way that current theories of measurement
aspire to merge with standard quantum mechanics, or make the neurological connection with conscious observation.

\section*{Another Rule-Set}
	Another rule-set called the \emph{nRules (1-4)} are given in detail in another \mbox{paper \cite{RM2}}.  These are similar to
the oRules except that they allow both an observer type measurement and an objective type measurement. The former type occurs only
in the presence of an observer, whereas the latter takes place independent of an observer.  These rules therefore come closer to
the spirit of traditional measurement theory than do the oRules, but they are still a significant departure because they also
introduce probability through the notion of `probability current' rather than square modulus \emph{and} because they address the
state reduction of conscious individuals  in an ontological context.

\section*{Purpose of Rule-Sets}
It is possible to have an empirical science using the epistemological model without explicitly talking about consciousness. 
This is because it is always assumed that the outside observer is conscious, so there is no need to make a theoretical
point of it.   
 
However, in the ontological model, everything that exists is in principle included in the system.  So if quantum mechanics is to
be an empirical science, then the system must provide for the existence of conscious brains that can make empirical
observations.  This means that the theory must be told when and how conscious brain states appear so that an empirical science
is possible within the model.  Special rules like one of the above rule-sets are required for this purpose. 

I emphasize again that these rule-sets are not alternative theories that seek to replace the statistical formalism of von
Neumann.  Each applies to individuals, and is like an empirical formula that requires a wider theoretical framework in order to be
understood -- a framework that is presently unknown.  I do not finally choose one of the rule-sets or propose an explanatory
theory.  I am only concerned with the ways in which state reduction might work in each case.

\section*{The Interaction: Particle and Detector}
		Before introducing an observer into this ontological model, consider an interacting particle and detector system by itself. 
These two objects are assumed to be initially independent and given by the equation 
\begin{equation}
\Phi(t) = exp(-iHt)\psi_i\otimes d_i
\end{equation}
where $\psi_i$ is the initial particle state and $d_i$ is the initial detector state.  The particle is then allowed to pass over
the detector, where the two interact with a cross section that may or may not result in the capture.  After the interaction
begins at a time $t_0$, the state is an entanglement in which the particle variables and the detector variables are not
separable.  

However, we let $\phi(t \ge t_0)$ be in a representation whose components can be grouped so that the first component includes the
detector $d_0$ in its ground state prior to capture, and the second component includes the detector $d_1$ in its capture state. 
There is then a clear discontinuity or ``quantum jump" between the two components.    The captured particle is included in $d_1$
in the second component, giving
\pagebreak
\begin{equation}
\Phi(t \ge t_0) = \psi(t)d_0 + d_1(t)
\end{equation}
where $d_1(t)$ is equal to zero at $t_0$ and increases with time\footnote{Each component in eq.\ 2 has an attached environmental
term $E_0$ and $E_1$.  These are orthogonal, insuring local decoherence. The equation appears to be a mixture because these terms
are not shown.  However, eq.\ 2 (including the environmental terms) and others like it are fully coherent superpositions, and
in the following we will call them ``superpositions", reflecting their global rather than their local
properties.}$^,$\footnote{Superpositions of environmentally isolated macroscopic states have been found at low
temperatures \cite{JRF}.  These superpositions are observed through interference effects between them.  Although no such
interference exists in eq.\ 2, we nonetheless assume that the  locally decoherent macroscopic states $d_0$ and $d_1$ are in global
superposition.}. 
$\psi(t)$ is a free particle as a function of time, including all the incoming and scattered components.  It does no harm and it
is convenient to let $\psi(t)$ carry the total time dependence of the first component, and to let $d_0$ be normalized
throughout\footnote{Equation  2 can be written with coefficients $c_0(t)$ and $c_1(t)$ giving $\Phi(t \ge t_0) =
c_0(t)\psi(t)d_0 + c_1(t)d_1$, where all three states $\psi(t)$, $d_0$, and $d_1$ are normalized throughout.  We let
$c_0(t)\psi(t)$ in this expression be equal to $\psi(t)$ in eq.\ 2, and let $c_1(t)d_1$ be equal to $d_1(t)$ in eq.\ 2.}. 

The first component in eq.\ 2 is a superposition of all possible scattered waves of $\psi(t)$ in product with all possible
recoil states of the ground state detector, so $d_0$ is a spread of states including all the recoil possibilities together with
their correlated environments.  The second component is also an entangled superposition of this kind.  This one includes all the
recoil components of the detector that have captured the particle.  
There will be no state reduction of \mbox{eq.\ 2} in this oRule treatment because there is no observer present.  The interaction
will continue until it is complete at a time $t_f$, after which time
\begin{displaymath}
\Phi(t \ge t_f > t_0) = \psi(t)d_0 + d_1(t_f)
\end{displaymath}

\section*{Add an Observer}
Assume that an observer is looking at the detector in eq.\ 1 from the beginning.  
\begin{displaymath}
\Phi(t ) = exp(-iHt)\psi(t)_i\otimes D_iB_i
\end{displaymath}
where $B_i$ is the initial brain state of the observer that is entangled with the detector.  This brain is understood to include
\emph{only} higher order brain parts \mbox{-- that is}, the physiology of the brain that is directly associated with consciousness
after all image processing is complete.  All lower order physiology leading to $B_i$ is assumed to be part of the detector.  The
detector is now represented by a capital $D$, indicating that it includes the bare detector by itself \emph{plus} the low-level
physiology of the observer.

Following the interaction between the particle and the detector we will have
\begin{equation}
\Phi(t \ge t_0) = \psi(t)D_0B_0 + D_1(t)B_1
\end{equation}
where $B_0$ is the observer's brain when the detector is observed to be in its ground state $D_0$, and $B_1$ is the brain state
when the detector is observed to be in its capture state $D_1$.   

If the interaction is long lived compared to the time it takes
for the detector to record the changes in eq.\ 3, then the superposition in that equation might exist for a long time
before a capture causes a state reduction.  This suggests that there are two active brain states of this observer that might be
simultaneously observing the detector, where one sees the detector in its ground state and the other sees it in its capture
state.  The equation therefore invites a paradoxical interpretation like that associated with Schr\"{o}dinger's cat.  This
ambiguity cannot be allowed.  The oRules of this paper must not only provide for a stochastic trigger that gives rise to a state
reduction, and describe that reduction, they must also insure than an empirical ambiguity of this kind does not exist.  

To this end we introduce dual brain state categories `realized' and `ready', where realized brain states are conscious, and may
be thought of as more ``real" than ready brain states that are \emph{not} conscious.  The latter are only on stand-by, ready to
be stochastically chosen and converted to conscious states after state reduction.

\section*{Ready Brains}
A \emph{realized brain state B} (not underlined) is assumed to be conscious of something with which it is entangled -- like
$B_0$ is aware of $D_0$ in eq.\ 3.  The corresponding ready brain state $\underline{B}$ (underlined) has the same content as its
partner $B$ except that it is not conscious.  That is not to say that $\underline{B}$ is \emph{un}conscious.  It is more like a
`potential' state of the conscious state $B$.  In the following, an \emph{active} brain state is defined to be one that is
actively engaged in an observation -- i.e., it is realized or ready but not \emph{un}conscious of the object in question.  There
are four symbols that may be used to represent brain states.

\rightskip=.5in
\hangindent=.5in
$B$ \emph{realized} brain state -- an active brain state that is understood to be conscious.

\hangindent=.5in
$\underline{B}$ \emph{ready} brain state -- an active brain state with the same  content 
as $B$, except that it is \emph{not} conscious.  Ready brain states  are always underlined.

\hangindent=.5in
$B^{b}$ \emph{brink} state -- an inactive brain state that is on the brink of  becoming an 
active brain state.  Inactive with respect to 
$B$ means:  neither $B$ nor $\underline{B}$.

$X$ \emph{unknown} brain state 

\rightskip=0in

\section*{The oRules}
The first rule establishes the existence of a stochastic trigger.  This is a property of the system that has nothing to do with
the kind of interaction taking place or its representation.  Apart from making a choice, the trigger by itself has no effect on
anything.  It initiates a state reduction only when it is combined with oRules 2 and 3.

\vspace{.4cm}

\noindent
\textbf{nRule (1)}: \emph{For any subsystem of n components in a system having a total square modulus equal to s, the
probability per unit time of a stochastic choice of one of those components at time t is given by $(\Sigma_nJ_n)/s$, where
the net probability current $J_n$ going into the $n^{th}$ component at that time is positive.}

\vspace{.4cm}

The second rule specifies the conditions under which ready brain states appear in solutions of Schr\"{o}dinger's equation. 
These are understood to be the basis states of a state reduction\footnote{The wording of oRule (2) and oRule (3) is slightly
different from the published wording of rule (2) and rule (3) in ref.\ 1.}.

\vspace{.4cm}

\noindent
\textbf{oRule (2)}: \emph{If an interaction produces new components that are discontinuous with the initial state or with each
other, then all of the  active brain states in the new components will be ready brain states.}

\noindent
[\textbf{note}: \emph{Continuous}  means, continuous in all variables. Although solutions to Schr\"{o}dinger's equation change
continuously in time, they can be discontinuous in other variables -- e.g., the separation between the $n^{th}$ and the $(n +
1)^{th}$ orbit of an atom with no orbits in between.  Of course atomic states are generally coherent, but a discontinuity of this
kind can also exist between macroscopic states that are decoherent.  For instance, the displaced detector states $D_0$ (ground
state) and $D_1$ (capture state) are discontinuous with respect to detector variables.  There is no eigenstate $D_{1/2}$  in
between.  Like atomic orbits, these two detector states are a `quantum jump' apart.]

\noindent
[\textbf{note}: The \emph{initial state} is the initial state of the system that appears in a given solution of Schr\"{o}dinger's
equation.  A particular solution is defined by a unique set of boundary conditions.  So eqs.\ 1 and 2 are both included in the
single solution that contains the discontinuity between $d_0$ and $d_1$, where eq.\ 1 is the initial state.  However, boundary
conditions change with the collapse of the wave function, so the component that survives a collapse becomes the initial state of
the new solution.]

\vspace{.4cm}

\noindent
\textbf{oRule (3)}: \emph{If a component containing ready brain states is stochastically chosen, then those states will become
realized (i.e., conscious) brain states, and all other components in the superposition will be immediately reduced to zero.}

\noindent
[\textbf{note}: The claim of an immediate (i.e., discontinuous) reduction is the simplest way of describing the collapse of the
state function.  The collapse is brought about by an instantaneous change in the boundary conditions of the Schr\"{o}dinger
equation, rather than by the introduction of a new `continuous' mechanism of some kind.  A continuous modification can be added
later (with a modification of oRule 3) if that is seen to be necessary\footnote{The new boundary comes from a stochastic hit on
one of the available eigenvalues, which \emph{is} the new boundary.  In this treatment the stochastic trigger is intrinsically
discontinuous and imposes that discontinuity on the developing wave function.  This is the simplest way to account for the
sudden change that takes place, and it spares our having to explain where a change producing continuous mechanism `comes
from'.}.]

\noindent
[\textbf{note}: This collapse does not generally preserve normalization.  That does not alter the probability in subsequent
reductions because of the way probability per unit time is defined in oRule (1) -- that is, divided by the total square modulus.]

\vspace{.4cm}

The fourth oRule forbids transitions from components containing ready states to other components.  Only positive current going
into ready states is physically meaningful because positive current represents a positive probability of reduction.    A negative
current (coming out of a ready brain state) is not physically meaningful, and is not allowed by oRule (4).  

\vspace{.4cm}

\noindent
\textbf{oRule (4)}: \emph{If a component in a superposition  is entangled with a ready brain
state, then that component can only receive probability current.}

\vspace{.4cm}

If an interaction does not produce ready brain components that are discontinuous with the given states or with each other, then
the Hamiltonian will develop the state in the usual way, independent of these rules.  If the stochastic trigger selects a component
that does not contain ready brain states, then there will be no oRule (3) state reduction.

\section*{Apply to Interaction}
When these rules are applied to eq.\ 3, we have 
\begin{equation}
\Phi(t \ge t_0) = \psi(t)D_0B_0 + D_1(t)\underline{B}_1
\end{equation}
where the brain state in $D_1(t)\underline{B}_1$ is a ready state by virtue of oRule (2), so it is not conscious.  Since there is
only one conscious brain state in this superposition, a cat-like ambiguity is avoided.  Equation 4 (with underline) now
\emph{replaces} eq.\ 3.

	Equation 4 is the state of the system before there is a stochastic hit that produces a state reduction.  The observer is here
consciously aware of the detector in its ground state $D_0$, for the brain state $B_0$ is correlated with $D_0$.  If there is a
capture, then there will be a stochastic hit on the second component in eq.\ 4 at a time $t_{sc}$.  This will reduce the first
component to zero according to oRule (3), and convert the ready state in eq.\ 4 into a conscious brain state.
\begin{equation}
\Phi(t \ge t_{sc} > t_0) = D_1(t)B_1
\end{equation}

Standard quantum mechanics (without these rules) gives us eq.\ 3 by the same logic that it gives us Schr\"{o}dinger's cat and
Everett's many worlds.  Equation 3 is a single equation that simultaneously presents two different conscious brain states, thus
assuring an unacceptable ambiguity.  However with these oRules in effect, the Schr\"{o}dinger solution is properly grounded in
observation, allowing the rules to correctly and unambiguously predict the experience of the observer.  This replaces
`one' equation eq.\ 3 with `two' equations in eqs.\ 4 \mbox{and 5}.   \mbox{Equation 4} describes the state of the system
\emph{before} capture, and eq.\ 5 describes the state of the system \emph{after} capture.  Before and after are two different
solutions to Schr\"{o}dinger's equation, specified by different boundary conditions.  Remember, we said that the stochastic
trigger selects the (new) boundary that applies to the reduced state.  So it is the stochastic event that separates the two
solutions  -- defining before and after.  

If there is no stochastic hit on the second component in eq.\ 4, then it will become a \emph{phantom} component.  A component is
a phantom when there is no longer any probability current flowing into it (in this case because the interaction is complete), and
when there can be no current flowing out of it because it is composed of ready states that comply with oRule (4).  A phantom
component can be dropped out of the equation without consequence.  Doing so only changes the definition of the system. It is the
same kind of redefinition that occurs in standard practice when one chooses to renormalize a system at some new starting time. 
Keeping a phantom is like keeping the initial system.  Because of nRule (3), kept phantoms are reduced to zero whenever another
component is stochastically chosen.  

	If there is no stochastic hit in eq.\ 4, then the new system (dropping the phantom $D_1\underline{B}_0$) is just the first
component of that equation.  This corresponds to the observer continuing to see the ground state detector $D_0$, as he should in
this case.

\section*{A Terminal Observation}
	An observer who is inside a system must be able to confirm the validity of the Born rule that is normally applied from the
outside.  To show this, suppose our observer is not aware of the detector during the interaction with the particle as in eq.\ 2,
but he looks at the detector \emph{after} the interaction is complete.  During the interaction we then have 
\begin{displaymath}
\Phi(t_f > t > t_0) = [\psi(t)d_0 + d_1(t)]\otimes X
\end{displaymath}
where $t_f$ is the time of completion, and $X$ is the unknown  state of the observer prior to the physiological
interaction\footnote{The ``decision" of the observer to look at the detector is assumed to be deterministically internal in the
ontological model.  In this respect, the ontological model is like classical physics.}. 

After the interaction is complete and before the observer looks at the detector 
\begin{displaymath}
\Phi(t \ge t_f > t_0) = [\psi(t)d_0 + d_1(t_f)]\otimes X
\end{displaymath}
where there is no longer a probability current flow inside the brackets.  When the observer finally looks at the detector at
time $t_{look}$, we have
\begin{eqnarray}
\Phi(t \ge t_{look} >t_f > t_0) &=& [\psi(t)d_0 +  d_1(t_f)]\otimes X\\
&\rightarrow& [\psi'(t)D_0 +  D_1(t_f)]B^b \nonumber
\end{eqnarray}
where the physiological process (represented by the arrow) carries $\otimes X$ into $B^b$, $d_0$ into $D_0$, and $d_1$ into
$D_1$ by a continuous classical progression leading from independence to entanglement.  The brain state $B^b$ is understood to
be an inactive state on the brink of becoming active. There are as yet no conscious states in eq.\ 6 because the process has not
gotten beyond the brink state -- i.e., all the brain states in eq.\ 6 are inactive with respect to the detector.  

During this process the observer will be unable to distinguish between the two detector states $D_0$ and $D_1$, which is why his
brain is called inactive at this time.  However, at the moment of observation $t_{ob}$, he will resolve the difference between
these states, and when that happens a continuous `classical' evolution will no longer be possible.  The solution will then branch
``quantum mechanically" into two components that separately recognize $D_0$ and $D_1$.  
\begin{eqnarray}
\Phi(t \ge t_{ob} > t_{look} >t_f > t_0) &=& \psi(t)D_0B^b +  D_1(t_f)B^b\\
&+& \psi'(t)D_0\underline{B}_0 +  D'_1(t_f)\underline{B}_1 \nonumber
\end{eqnarray}
where the components in the second row are zero at $t_{ob}$ and increase in time. Current flows vertically during this active
phase of the physiological interaction.  The states in the second row are discontinuous from each other (i.e., $D_0$ and $D_1$
are discontinuous) and contain active brain states.  They are therefore required by oRule (2) to be ready states.  It is here
that the non-conscious nature of ready states is important.  Otherwise, eq.\ 7 would give us an ambiguous dual conscious
(cat-like) result.  

	With probability current flowing into the second row of eq.\ 7, there is a probability equal to 1.0 that one of those components
will be stochastically chosen.  If the third component is chosen at a time $t_{sc3}$, then oRule (3) will give
\begin{equation}
\Phi(t \ge t_{sc3} > t_{ob} > t_{look} > t_f > t_0) = \psi(t)D_0B_0
\end{equation}
indicating that the terminal observer finds that the particle was not captured during the primary interaction.  

If the fourth component is chosen at a time $t_{sc4}$, then oRule (3) will give 
\begin{equation}
\Phi(t \ge t_{sc4} > t_{ob} > t_{look} > t_f > t_0) = D_1(t_{sc4})B_1
\end{equation}
indicating that the terminal observer finds that the particle \emph{was} captured during the primary interaction.  The probability
of eq.\ 8 plus eq.\ 9 is equal to 1.0, thereby confirming the Born interpretation.

\section*{An Intermediate Case}
		In eq.\ 4 the observer is assumed to interact with the detector from the beginning.  Suppose that the incoming particle results
from a long half-life decay, and that the observer's physiological involvement only \emph{begins} in the middle of that
interaction.  Before that time we have
\begin{displaymath}
\Phi(t \ge  t_0) = [\psi(t)d_0 + d_1(t)]\otimes X
\end{displaymath}
where again $X$ is the unknown brain state of the observer prior to the physiological interaction.  Primary probability current
here flows between the detector components inside the bracket.    

Let the observer interact with the detector at some time $t_{look}$ giving
\begin{eqnarray}
\Phi(t \ge t_{look} > t_0) &=& [\psi(t)d_0 +  d_1(t)]\otimes \nonumber X\\
&\rightarrow& [\psi'(t)D_0 +  D_1(t)]B^b \nonumber
\end{eqnarray}
where the physiological process (represented by the arrow) carries $\otimes X$ into $B_b$, $d_0$ into $D_0$, and $d_1$ into
$D_1$ by a continuous classical progression leading from independence to entanglement.  The state $B^b$ is again understood to
be an inactive brain state on the brink of becoming active.  As before, the observer will be unable to distinguish
between the two detector states $D_0$ and $D_1$ during this process.  A resolution occurs at time $t_{ob}$ leading to  
\begin{eqnarray}
\Phi(t \ge t_{ob} > t_{look} > t_0) &=& \psi(t)D_0B^b +  D_1(t)B^b  \\
&+& \psi'(t)D_0\underline{B}_0 +  D'_1(t)\underline{B}_1 \nonumber
\end{eqnarray}
where the ready brain components in the second row are zero at $t_{ob}$ and increase in time.  Probability current flows
vertically into those components during the active phase of the physiological interaction.  Primary current flows horizontally
in the first row but not between the ready components in the second row because  $\underline{B}_1$ cannot receive current from
$\underline{B}_0$ according to oRule (4).  

	All of the current from the first component in eq.\ 10 will either collect in the third component or in the fourth component via
the second component.  The significance of oRule (4) in this case is that once probability is assigned to the third component,
it cannot be passed along to the fourth component.  The significance of the non-conscious nature of the ready states is the same
as it is in eq.\ 7 -- i.e., that the second row in eq.\ 10 will not give ambiguous results.    

If the vertical current going into the fourth component $D'_1(t)\underline{B}_1$ of eq.\ 10 results in a stochastic hit at time
$t_{sc4}$, the resulting state reduction will be   
\begin{equation}
\Phi(t \ge t_{sc4} > t_{ob} > t_{look} >  t_0) = D_1(t_{sc4})B_1
\end{equation}
indicating that the capture had already occurred by the time of the observation.  We said that the primary interaction is still in
progress when the observer looks at the detector.  This means that an observation may reveal a prior capture, even though
the actual reduction does not occur (in these oRules) until the observer makes an observation.  

	If the current going into the third component $\psi'(t)D_0\underline{B}_0$ of eq. 10 gives rise to a stochastic hit at time
$t_{sc3}$, the resulting state reduction will be
\begin{equation}
\Phi(t \ge t_{sc3} > t_{ob} > t_{look} >  t_0) = \psi'(t)D_0B_0 + D_1(t)\underline{B}_1
\end{equation}
where the second component is zero at $t_{sc3}$ and increases in time because the primary interaction is still going on.  If there
is stochastic hit on this  component  at a later time $t_{sc32} > t_{sc3}$, then there will be a further reduction
giving
\begin{equation}
\Phi(t \ge t_{sc32} > t_{sc3} > t_{ob} > t_{look} >  t_0) = D_1(t_{sc32})B_1
\end{equation}
indicating that the observer first came on board at $t_{sc3}$ and found that the capture had not yet occurred.  Then he witnessed
the capture at $t_{sc32}$.

	If the primary interaction in eq.\ 12 runs out before there is a stochastic hit on the second component, then this equation will
go unchanged, except that the time dependence of $D_1(t)$ will be removed when the interaction is complete at time $t_f$ giving
$D_1(t_f)$.  In this case, the second component
$D_1(t_f)\underline{B}_1$ will become a phantom and can be ignored.

\section*{A Second Observer}
If a second observer is standing by while the first observer interacts with the detector, the state function will be
\begin{displaymath}
\Phi(t \ge t_0) = [\psi(t)D_0B_0 + D_1(t)\underline{B}_1]\otimes X
\end{displaymath}
where $X$ is an unknown state of the second observer prior to his interacting with the system.  The detector $D$ here includes the
low-level physiology of the first observer.  When a product of brain states appears in the form $BB$ or $B\otimes X$, the first
term will refer to the first observer and the second to the second observer.  

If the second observer interacts with the detector at time $t_{ob}$ (skipping $t_{look}$) the result of the physiological
interaction would seem to be
\begin{eqnarray}
\Phi(t \ge t_{ob} > t_0) &=& \psi(t)D_0B_0B^b +  D_1(t)\underline{B}_1B^b  \\
&+& \psi'(t)D_0B_0\underline{B}_0 +  D'_1(t)\underline{B}_1\underline{B}_1 \nonumber
\end{eqnarray}
where the second row follows from the active physiological interaction and \mbox{oRule (2)}.  A further expansion of the detector
is assumed to include the low-level physiology of the second observer.

However, the fourth component of eq.\ 14 will remain zero amplitude because oRule (4) forbids a physiological current to flow
into it from either the second or third component.  In addition, the third component evolves continuously from the first
component.  Therefore, since there is no other product of the psychological interaction that is discontinuous with it, the third
component is not required by oRule (2) to be a ready state. So beginning with $t_{look}$, the solution generated by the
physiological interaction and going through to just past $t_{ob}$ will carry the first component of eq.\ 14 into
$\psi(t)D_0B_0B_0$ by a continuous classical process giving 
\begin{eqnarray}
\Phi &=& \psi(t_{look})D_0B_0\otimes X \rightarrow  \psi(t_i)D_0B_0B^i \rightarrow \psi(t_{ob+})D_0B_0B_0  \\
&+& D_1(t_{look})\underline{B}_1\otimes X \rightarrow  D_1(t_i)\underline{B}_1B^i \rightarrow
D_1(t_{ob+})\underline{B}_1\underline{B}_1\nonumber
\end{eqnarray}
\emph{instead} of eq.\ 14.  The index $i$ refers to any time between $t_{look}$ and $t_{ob}$ when the inactive brain state of the
second observer is $B^i$.  The first row of this equation is the single component that carries the observer all the way from
independence to entanglement. The arrows represent the brain going continuously from $X$ \mbox{to $B_0$}.  

The primary interaction is still active during this time, and this gives rise to a vertical current going from the first to the
second row in eq.\ 15.  The  second row is therefore a continuum of components that are created parallel to
the first row at each moment of time.  So at the time $t_{ob+}$, vertical current flows only into the final component in the
second row of eq.\ 15.  Components prior the last one no longer have current flowing into them from above, and since there is no
horizontal current among these ready states, they become phantom states as soon as they are created.  Therefore, when the
interaction is complete after $t_{ob+}$ eq.\ 15 is
\begin{equation}
\Phi(t \ge t_{ob+} > t_{look} > t_0) = \psi(t)D_0B_0B_0 + D_1(t)\underline{B}_1\underline{B}_1
\end{equation}
where $D_1(t)\underline{B}_1\underline{B}_1$ is zero at $t_{ob}$ and increases in time.  This equation is the same as eq.\ 4
except that it includes two observers; so from this point on, it is as though both observers have been on board from the
beginning.  

	Suppose the primary interaction terminates before the second observer is fully on board.  Let it terminate at time $t_i$.  Then
there will be no further evolution after that time in the second row of eq.\ 15; so the entire second row will be made up of
phantom components.  Following time $t_{ob}$ we will then have just
\begin{equation}
\Phi(t \ge t_{ob+} > t_{look} > t_0) = \psi(t)D_0B_0B_0 
\end{equation}
This says that both observers experience the detector in its ground state at the end of the interaction.  There was no capture
before time $t_i$; and since the interaction is terminated, there is no further prospect of a capture.      

	Instead of terminating the interaction at time $t_i$, suppose there is a stochastic hit at that time.  In that case, the second
component in the second row of \mbox{eq.\ 15} will be chosen to give
\begin{displaymath}
\Phi(t = t_i > t_{look} > t_0) = D_1(t_i)B_1B^i 
\end{displaymath}
The inactive state $B^i$ will then continue its evolution so that we finally have 
\begin{equation}
\Phi(t \ge t_i > t_{look} > t_0) = D_1(t_i)B_1B^i \rightarrow  D_1(t_i)B_1B_1
\end{equation}

\section*{Anomaly Avoided}
The fourth oRule avoids a catastrophic anomaly if the primary interaction is complete at time $t_f$ \emph{without} a capture, and
before the second observer looks at the detector.  
\begin{displaymath}
\Phi(t \ge t_{f} > t_0) = [\psi(t)D_0B_0 + D_1(t_f)\underline{B}_1]\otimes X
\end{displaymath}
After the second observer observes the detector at $t_{ob}$ we will have
\begin{eqnarray}
\Phi(t \ge t_{ob} > t_f > t_0) &=& \psi(t)D_0B_0B^b +  D_1(t_f)\underline{B}_1B^b  \\
&+& \psi'(t)D_0B_0\underline{B}_1 + D'_1(t_f)\underline{B}_1\underline{B}_1 \nonumber
\end{eqnarray}
where the second row is zero at $t_{ob}$ and increases in time.  This differs from eq.\ 14 only in that the primary interaction is
already complete.  There is no horizontal current flow.

Assume that oRule (4) is not in effect.  In that case the fourth component $D'_1(t_f)\underline{B}_1\underline{B}_1$ in eq.\ 19
would be accessible to current from the second component.  A stochastic hit at some time $t_{sc4}$ would then be possible,
yielding
\begin{displaymath}
\Phi(t \ge t_{sc4} > t_{ob} > t_{f} > t_0) =  D_1(t_f)B_1
B_1\end{displaymath}
This says that even though the first observer can testify that the interaction has been completed without a capture, both
observers will experience a capture when the second observer comes on board -- some time \emph{after} the interaction is
completed.  That is absurd.  The fourth oRule therefore plays the essential role in preventing absurdities of this kind.

\section*{A Counter}
	In the previous sections we have seen how the oRules go about including observers inside a system in an ontological model.  The
rules describe when and how the observer becomes conscious of measuring instruments; and furthermore, they replicate common
empirical experience in these situations.  In the next few sections we turn attention to another problem -- the requirement that
observed macroscopic states must appear in their normal sequence.   This sequencing chore represents a major application of oRule
(4) that is best illustrated in the case of a macroscopic counter.

\vspace{0.4cm}

Consider a $\beta$ counter in the ontological model where an observer interacts with the counter.  If the counter is turned on at
time $t_0$, its state function will be given by
\begin{equation}
\Phi(t \ge   t_0) =C_0(t)B_0 + C_1(t)\underline{B}_1
\end{equation}
where $C_1(t)\underline{B}_1$ is zero at $t_0$ and increases in time.  $C_0$ is a counter that reads zero counts, and $B_0$ is an
entangled conscious brain state that experiences the counter reading zero counts.   $C_1$ reads one count, and $C_2$ (not shown)
reads two counts, etc.  The ready brain state $\underline{B}_1$ appears as required by oRule (2).  

Components beyond $C_1(t)\underline{B}_1$ do not appear in eq.\ 20 because current cannot leave that component according to oRule
(4).  Therefore, the $0^{th}$ and the $1^{st}$ components are the only ones that are actively involved before there is a
stochastic hit of any kind.  Before that time, the only current flow will be $J_{01}$ from the $0^{th}$ to the $1^{st}$
component.  The resulting distribution at some time $t < t_{sc}$ is shown in fig.\ 3, where $t_{sc}$ is the time of a stochastic
hit on the second component.

\begin{figure}[b]
\centering
\includegraphics[scale=0.8]{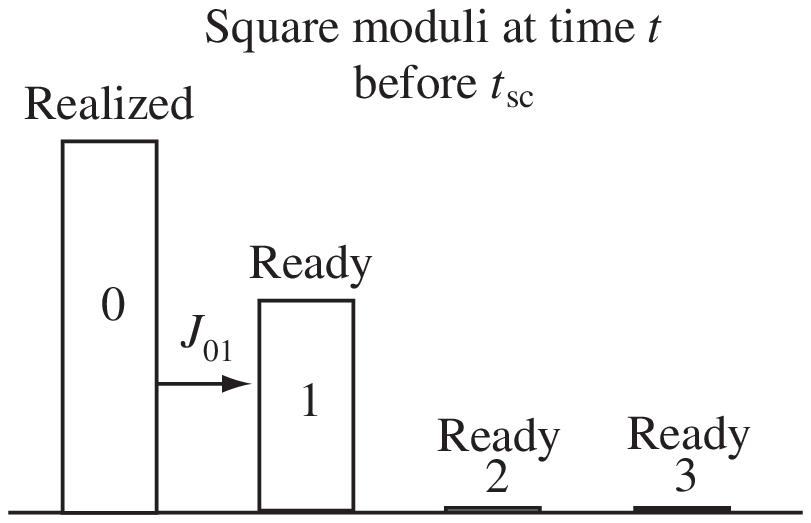}
\center{Figure 3}
\end{figure}

This means that because of oRule (4), the $1^{st}$ component \emph{will} be chosen because all of the current from the (say,
normalized) $0^{th}$ component will pore into the $1^{st}$ component making $\int J_{01}dt = 1.0$.  The first two dial readings
will therefore be sequential, going from 0 to 1 without skipping a step such as going from 0 to 2.  

Unobserved macroscopic counters will develop as a superposition of states (see next section), allowing the first stochastic hit to
occur on any component $C_1$, $C_2$, $C_3$, etc.  However, oRule (4) enforces a no-skip behavior of macroscopic objects; so
observed macroscopic things will always follow in sequence without skipping a step. 

With the stochastic choice of the $1^{st}$ component at $t_{sc}$, the process will begin again as shown in fig.\ 4.  This also
leads with certainty to a stochastic choice of the $2^{nd}$ component.  That certainty is accomplished by the wording of
\mbox{oRule (1)} which requires that the probability per unit time is given by the current flow $J_{12}$ divided by the total
square modulus at that moment.  The total integral $\int J_{12}dt$  is less than 1.0 in fig.\ 4, but it is restored to 1.0 when
divided by the total square modulus.  It is therefore certain that the $2^{nd}$ component will be chosen.

\begin{figure}[h]
\centering
\includegraphics[scale=1.1]{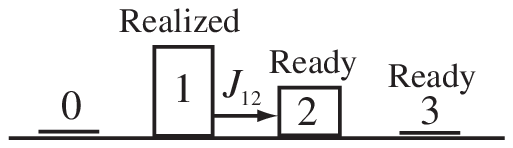}
\center{Figure 4}
\end{figure}

And finally, with the choice of the $2^{nd}$ component, the process will resume again with current $J_{23}$ going from the
$2^{nd}$ to the $3^{rd}$ component.  This leads with certainty to a stochastic choice of the $3^{rd}$ component.

\section*{The Counter with Delayed Observation}
When the observer is not observing the counter, eq.\ 20 is written
\begin{equation}
\Phi(t \ge   t_0) = [C_0(t) + C_1(t) + C_2(t) + C_3(t) + . . . \hspace{.1cm} etc.] \otimes X 
\end{equation}
where all the components following $C_0$ are zero at $t_0$.  $X$ is the unknown state of the independent observer.  Immediately
after $t_0$, current $J_{01}$ flows from the $0^{th}$ component to the $1^{st}$ component, but not to higher order components
because the Hamiltonian only connects the $0^{th}$ with the $1^{st}$.  However, current $J_{12}$ will begin to flow into the
$2^{nd}$ component as soon as the 1st acquires a non-zero amplitude.  The $3^{rd}$ component will also receive current $J_{23}$
when the $2^{nd}$ acquires amplitude; so after a time $t$, the distribution might look like fig.\ 5.

\begin{figure}[h]
\centering
\includegraphics[scale=0.8]{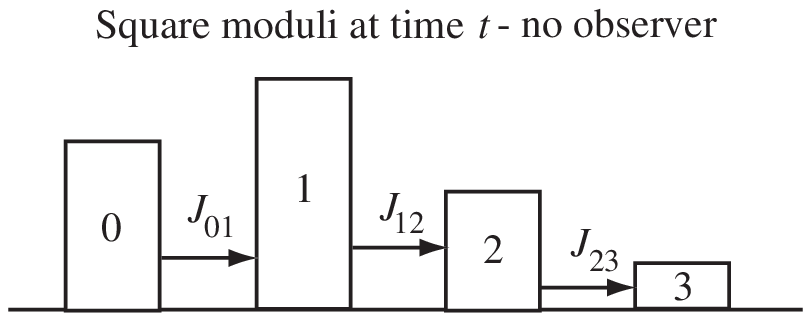}
\center{Figure 5}
\end{figure}

\pagebreak

Without an observer the macroscopic detector behaves like a familiar quantum mechanical object.  It will be a superposition of
many possibilities.  If the observer then interacts with the counter at time $t_{ob}$, oRule (2) requires that \mbox{eq.\ 21}
becomes 
\begin{eqnarray}
 \Phi(t \ge t_{ob} > t_0) &=& C_0(t)B^b + C_1(t)B^b + C_2(t)B^b + C_3(t)B^b + \hspace{.1cm}  \\
&+&  C_0(t)\underline{B}_0 + C_1(t)\underline{B}_1 + C_2(t)\underline{B}_2 + C_3(t)\underline{B}_3 + \hspace{.1cm}
\nonumber
\end{eqnarray}
where the second row is zero at $t_{ob}$.  Again, physiological current flows down.  Horizontal current cannot flow in the second
row because of oRule (4).  If a stochastic hit occurs at time $t_{sc4}$ on the fourth component in the second row of eq.\ 22, we
will have
\begin{equation}
\Phi(t = t_{sc4} > t_{ob} >   t_0) = C_3(t_{sc4})B_3 
\end{equation}
The state reduction in eq.\ 23 occurs with a probability that reflects the square modulus of the component $C_3$ in eq.\ 21.  The
Born rule is therefore verified in this application of the oRules.  Both the non-conscious property of a ready brain state and its
oRule (4) property are put to use in eq.\ 22.  

	Suppose the observer looks at the counter at the initial time $t_0$ and then leaves the room.  While he waits in the hall outside
his lab, the counter will evolve as a quantum mechanical superposition of states like those in fig.\ 5.  When he returns to look
at the counter, he will see just one result as in eq.\ 23.  So far as he is concerned, the counter behaved in an entirely classical
way while he was in the hall.  He will not know if the system follows the oRules or the Born rule of standard quantum mechanics. 
Furthermore, there is no experiment that he can perform that will tell the difference.

\section*{A Film Record}
Suppose we try to determine what happened in the absence of the observer by taking a motion picture of the counter during that
time.  In that case the film in the camera would also evolve quantum mechanically, so every component $C_m(t)$ will have a
film-strip correlated with it that is made up of separate frames.  Each frame is designated by the letter $F$, so the state
equation after $t_0$ will be
\begin{eqnarray}
 \Phi(t \ge t_0) &=& F _0F.F.F.F.C_0(t) + F_0F_1F.F.F.C_1(t) + \nonumber\\
&+&  F _0F_1F_2F.F.C_2(t) + F _0F_1F_2F_3F.C_3(t) + ... \nonumber
\end{eqnarray}
where only the first component is non-zero at $t_0$.  The symbol $F.$ (with a dot) refers to a film frame that is not yet
exposed.    In the first component, only the first frame is exposed to the counter state $C_0$ and the remaining frames are as yet
unexposed.  The camera is arranged so that a new frame is exposed as soon as a new count is registered.  So in the second
component the first and second frames are exposed when the second is exposed to the counter state $C_1$.  The remaining frames in
that component are unexposed.

To simplify the notation, let $F_0F_1F_2F_3F.C_3(t)$ be written $F_{-3}..C_3(t)$ where the      sub-dash represents all the
numbers before the number 3.  The equation is then
\begin{displaymath}
\Phi(t \ge t_0) = [F_0..C_0(t) + F_{-1}..C_1(t) + F_{-2}..C_2(t) + F_{-3}..C_3(t) + ...]\otimes X
\end{displaymath}
where the observer is shown waiting in the hall.  When the observer enters the room and observes the counter at time $t_{ob}$, the
interaction will yield 
\begin{eqnarray}
 \Phi(t \ge t_{ob} > t_0) &=& F_0..C_0(t)B^b + F_{-1}..C_1(t)B^b + F_{-2}..C_2(t)B^b + ... \nonumber\\
&+&  F_0..C_0(t)\underline{B}_0 + F_{-1}..C_1(t)\underline{B}_1 + F_{-2}..C_2(t)\underline{B}_2 + ... \nonumber
\end{eqnarray}
where the second row is zero at $t_{ob}$ and increases in time.  A stochastic hit on the third component $\underline{B}_2$ in
the second row at time $t_{sc3}$ gives a reduction similar to eq.\ 23
\begin{displaymath}
\Phi(t = t_{sc3} > t_{ob}) = F_{-2}..C_2(t_{sc3})B_2
\end{displaymath}
This could mean that the observer has become directly aware of either the counter reading $C_2$, or the third frame of the film
strip.  It doesnÕt matter since the two are correlated.

Let's suppose that after his observation of $F_{-2}..C_2(t_{sc3})B_2$, the observer looks at the first frame at time $t_{ob1}$ to
insure that it still reads 0 as he observed before leaving the room.  This will not require a stochastic trigger, for it involves
a purely classical inspection of the film strip that leads to  
\begin{displaymath}
\Phi(t \ge t_{ob1} > t_{sc3} > t_{ob}) = F_{-2}..C_2(t)B_2 \rightarrow F_{-2}..C'_2(t)B_{02}
\end{displaymath}
where the realized brain state $B_{02}$ is now conscious of both the 0 reading on the first frame and the 2 reading in the third
frame.  Continuing the investigation, the observer checks the second frame at time $t_{ob2}$. This also involves a classical
progression.
\begin{displaymath}
\Phi(t \ge t_{ob2} > t_{ob1} > t_{sc3} > t_{ob}) = F_{-2}..C'_2(t)B_{02} \rightarrow F_{-2}..C''_2(t)B_{012}
\end{displaymath}
where the realized brain state $B_{012}$ is conscious of the 0 reading on the first frame, the 1 reading on the second frame,  and
the 2 reading in the third frame.  There is only one stochastic occurrence in this case, assuming the counter was turned off
when the observer came back into the room.  

	Even though the state reduction can only be accomplished in the presence of an observer, the results can be verified by other
means (the film strip) that may or may not be immediately observed following the reduction.   Non-local correlations insure that
there will be complete consistency with all post-reduction investigations.  So far as the observer is concerned,  the detector
and camera behaved like classical instruments in his absence.  As a result of his observation and subsequent investigation, the
observer is justified in believing that the apparatus did not develop as a quantum mechanical superposition when he was in the
hall.

\section*{The Parallel Case}
Now imagine parallel states in which a quantum process may go either clockwise or counterclockwise as shown in fig.\ 6. 
Each component includes a macroscopic piece of laboratory apparatus A, where the Hamiltonian provides for a clockwise
interaction going from the $0^{th}$ to the $r^{th}$ state and from there to the final \mbox{state $f$}; as well as a
counterclockwise interaction from the $0^{th}$ to the $l^{th}$ state and from there to the final state $f$.  The Hamiltonian does
not provide a direct route from the $0^{th}$ to the final state. 

After being turned on at time $t_0$, the apparatus becomes a superposition 
\begin{displaymath}
\Phi(t \ge t_0) = A_0(t) + A_l(t) + A_r(t) + A_f(t)
\end{displaymath}
where the components following $A_0$ are zero at $t_0$. The state $A_0$ will then send current into $A_l$ and $A_r$, which in turn
send current to $A_f$.  A superposition will develop along these lines until the interaction ends.  There will be no stochastic
choice or state reduction because there is no observer present.
\begin{figure}[h]
\centering
\includegraphics[scale=0.8]{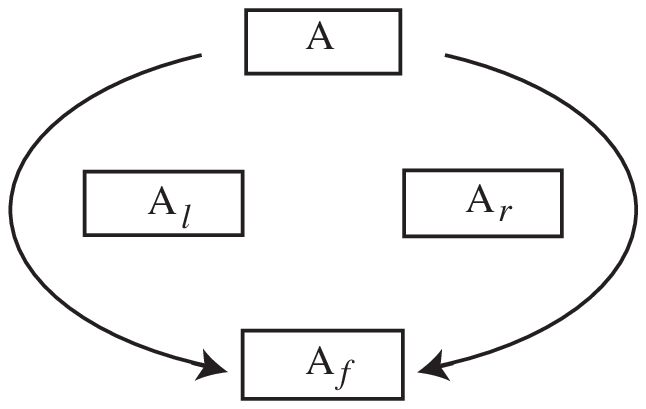}
\center{Figure 6}
\end{figure}

When the apparatus is being observed, the state will be
\begin{displaymath}
\Phi(t \ge t_0) = A_0(t)B_0 + A_l(t)\underline{B}_l + A_r(t)\underline{B}_r 
\end{displaymath}
where both the second and third components receive current directly from $A_0B_0$.  

With nRule (4) in place, probability current cannot initially flow from either of the intermediate states to the final
state, for that would carry a ready state into another state.  The dashed lines in fig.\ 7 indicate the forbidden
transitions.  But once the state $\underline{A}_l$ (or $\underline{A}_r$) has been stochastically chosen, it will become a
realized state $A_l$ (or $A_r$) and a subsequent transition to $\underline{A}_f$ can occur that realizes $A_f$.   
\begin{figure}[h]
\centering
\includegraphics[scale=1.0]{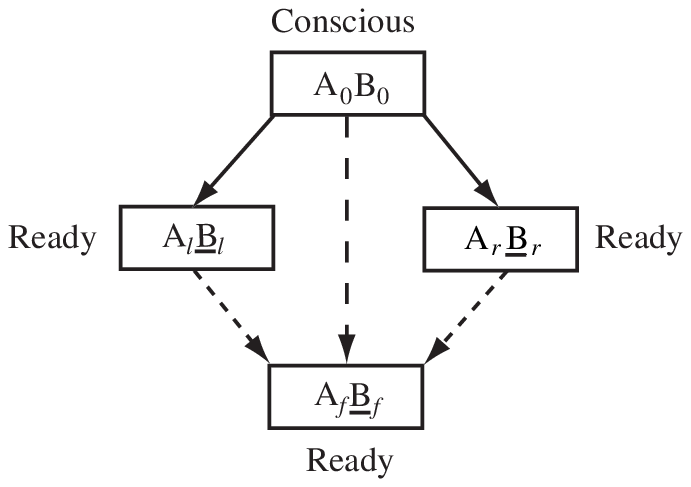}
\center{Figure 7}
\end{figure}

The effect of oRule (4) is therefore to force the system into a classical sequence that goes either clockwise or
counterclockwise.  Without it, the system might make a direct second order transition (through one of the intermediate states) to
the final state, without the intermediate states being realized.  The observer would then see the initial state followed by the
final state, without knowing which pathway was followed.  This is familiar behavior when the system is microscopic\footnote{In
Heisenberg's famous example formalized by Feynman, a microscopic particle observed at point $A$ and later at point $B$ will travel
over a quantum mechanical superposition of all possible paths in between.  Without nRule (4), macroscopic objects would do the
same thing.}, but it should not be the case in macroscopic systems.  The fourth oRule therefore forces the system into one or the
other classical path, so it is not a quantum mechanical superposition of both paths.

\section*{A Continuous Variable}
		In the above examples an observer plus oRule (4) guarantees that sequential steps in the macroscopic system are not passed
over.  If the variable itself is classical and continuous, then continuous observation is possible without the necessity of
stochastic jumps.  In that case we do not need oRule (4) or any of the oRules (1-4), for they do not prevent or in any way qualify
the motion.  

However, a classical variable may require a quantum mechanical jump-start.  For instance, the mechanical device that is used to
seal the fate of Schr\"{o}dinger's cat (e.g., a falling hammer) begins its motion with a stochastic hit.  That is, the decision to
begin the motion (or not) is left to a $\beta$-decay.  In this case, the presence of an observer (looking at the hammer) forces the
motion to begin at the beginning, insuring that no value of the classical variable is passed over; so the the hammer will fall
from its \emph{initial} angle with the horizontal.   Without oRule (4), or the observer, the hammer might begin its fall at some
other angle because probability current will flow into angles other than the initial one.  With an observer plus oRule (4) in
place, no angle will be passed over\footnote{In another paper \cite{RM6} the oRules are applied to two different versions of
the Schr\"{o}dinger cat experiment.}.

\section*{Grounding the Schr\"{o}dinger Solutions}
	Traditional quantum mechanics is not completely grounded in observation inasmuch as it does not include an observer.  The
epistemological approach of Copenhagen does not give the observer a role that is sufficient for him to realize the full empirical
potential of the theory; and as a result, this model encourages bizarre speculations such as the many-world interpretation of
Everett or the cat paradox of Schr\"{o}dinger.  However, when rules are written that allow a conscious observer to be given an
ontologically complete role in the system, these empirical distortions disappear.  It is only because of the incompleteness of the
epistemological model by itself that these fanciful excursions seem plausible\footnote{Physical theory should be made to
accommodate the phenomena, not the other way around. Everett goes the other way around when he creates imaginary phenomenon to
accommodate traditional quantum mechanics.  If the oRules were adopted in place of the Born rule, these flights of fantasy would
not be possible.}.

\section*{Limitation of the Born Rule}
		Using the Born rule, the observer can only record an observation at a given instant of time, and he must do so consistently over
an ensemble of observations.  He cannot himself become part of the system for any finite period of time.  He cannot become
continuously involved.  When discussing the Zeno effect it is said that continuous observation can be simulated by rapidly
increasing the number of instantaneous observations; but of course, that is not really continuous.    

	 On the other hand, the observer in an ontological model can \emph{only} be continuously involved with the observed system. 
That's because it takes a finite amount of time for the flow of physiological current to bring the observer on board to full
consciousness.  So the epistemological observer makes instantaneous observations but cannot make continuous ones; and the
ontological observer makes continuous observations but cannot make instantaneous ones.  Evidently the Born rule would requires
the ontological observer to do something that cannot be realistically done.  Epistemologically we can ignore this difficulty, but
a consistent ontology should not match a continuous physical process with continuous observation by using a discontinuous rule of
correspondence.  Therefore, an ontological model should not employ the Born interpretation that places unrealistic demands on an
observer.

\section*{Status of the Rules}
No attempt has been made to relate conscious brain states to particular neurological configurations.  The oRules are an
empirically discovered set of macrorelationships that exist on another level than micro-physiology, and there is no need to
connect these two domains.  These rules preside over physiological detail in the same way that thermodynamics presides over
molecular detail.  It is desirable to eventually connect these domains as thermodynamics is now connected to molecular motion; and
hopefully, this is what a covering theory will do.  But for the present we are left to investigate the rules by themselves without
the benefit of a wider theoretical understanding of state reduction or of conscious systems.  There are two rule-sets of this
kind, the oRules of this paper plus the nRules in ref.\ 4.  

	The question is, which of these two rule-sets is correct (or most correct)?  Without the availability of a wider theoretical
structure or a discriminating observation, there is no way to tell.  Reduction theories that are currently being considered may
accommodate a conscious observer, but none are fully accepted.  So the search goes on for an
extension of quantum mechanics that is sufficiently comprehensive to cover the collapse associated with an individual
measurement.  I expect that any such theory will support one of the ontological rule-sets, so these rules might serve as a guide
for the construction of a wider theory.

\end{document}